\documentstyle[11pt]{article}
\begin{document}

\def\b#1{{\bf #1}}
\def\i#1{{\it #1}}
\def\prob#1{{\bf P}\{#1\}}
\def\e#1{{\rm e}^{#1}}

\title{Sampling from a couple of positively correlated binomial variates}
\author{Mario Catalani\\
Department of Economics, University of Torino, Italy\\
mario.catalani@unito.it}
\date{}
\maketitle

\section{Introduction}
Let the random variables $\{X_1,\,X_2,\,X_3,\,X_4\}$ possess a multinomial
distribution ${\cal MN}(n,\,p_1,\,p_2,\,p_3,\,p_4)$
(see \cite{fishman}, \cite{kotz}). Just to fix notation
in this case the joint probability function is
\begin{eqnarray*}
\prob{X_1=x_i,\,\ldots,\,X_4=x_4}&=&P(x_1,\,\ldots,\,x_4)\\
&=&n!\prod_{i=1}^4{p_i^{x_i}\over x_i!},
\end{eqnarray*}
where $\sum_{i=1}^4p_i=1$, $\{x_i\}$ are integers $\ge 0$ and
$\sum_{i=1}^4X_i=n$.

\noindent
Now let $\{i_j\},\,j=1,\ldots,4$ define a permutation of the numbers
$\{1,\,2,\,3,\,4\}$ and define
$W=X_{i_1}+X_{i_2}$. Then
\begin{eqnarray}
\label{eq:fondamentale}
&&\prob{W=k,\,X_{i_3}=x_{i_3},\,X_{i_4}=n-k-x_{i_3}}\nonumber\\
&&\quad\quad =\sum_{i=0}^k\prob{X_{i_1}=i,\,X_{i_2}=k-i,\,
X_{i_3}=x_{i_3},\,X_{i_4}=n-k-x_{i_3}}\nonumber\\
&&\quad\quad =
\sum_{i=0}^k
{n!\over i!(k-i)!x_{i_3}!(n-k-x_{i_3})!}p_{i_1}^ip_{i_2}^{k-i}
p_{i_3}^{x_{i_3}}p_{i_4}^{n-k-x_{i_3}}\nonumber\\
&&\quad\quad =
{n!p_{i_3}^{x_{i_3}}p_{i_4}^{n-k-x_{i_3}}\over x_{i_3}!
(n-k-x_{i_3})!k!}\sum_{i=0}^k
{k!\over i!(k-i)!}p_{i_1}^ip_{i_2}^{k-i}\nonumber\\
&&\quad \quad = {n!p_{i_3}^{x_3}p_{i_4}^{n-k-x_{i_3}}
\over x_{i_3}!(n-k-x_{i_3})!k!}(p_{i_1}+p_{i_2})^k,
\end{eqnarray}
because of the binomial theorem.

\noindent
It follows $\{W,\,X_{i_3},\,X_{i_4}\}\sim
{\cal MN}(n,\,p_{i_1}+p_{i_2},\,p_{i_3},\,p_{i_4})$.
Because the marginals in a multinomial distribution are binomial variables
with the appropriate parameters we can conclude
\begin{equation}
\label{eq:fondamentaledue}
W=X_{i_1}+X_{i_2}\sim {\cal B}(n,\, p_{i_1}+p_{i_2}).
\end{equation}
Using this result define
$$\left\{\begin{array}{ccc}Y_1&=&X_1+X_3,\\
Y_2&=&X_2+X_3.\end{array}\right .$$
Then $Y_1\sim {\cal B}(n,\,p_1+p_3)$, $Y_2\sim {\cal B}(n,\,p_2+p_3)$.
Because in a multinomial distribution $\b{C}ov(X_i,\,X_j)=-np_ip_j$
we have
\begin{eqnarray*}
\b{C}ov(Y_1,\,Y_2)&=&-np_1p_2 -np_1p_3 -np_2p_3+np_3(1-p_3)\\
&=&-np_1p_2+np_3(1-p_1-p_2-p_3)\\
&=&-np_1p_2+np_3p_4\\
&=&n(p_3p_4-p_1p_2),
\end{eqnarray*}
so we have a positive covariance if $p_3p_4-p_1p_2>0$.

\noindent
Finally the linear correlation coefficient between $Y_1$ and $Y_2$ is given
by
$$\rho={p_3p_4-p_1p_2\over\sqrt{(p_1+p_3)(1-p_1-p_3)(p_2+p_3)(1-p_2-p_3)}}.$$

\bigskip
\noindent
Now suppose we want to sample from two given binomial variables with a
given positive correlation coefficient $r$. Let the given variables be
$Y_1\sim {\cal B}(n,\,\pi_1)$ and $Y_2\sim {\cal B}(n,\,\pi_2)$. So
we are assuming given the parameters $\pi_1,\,\pi_2,\,r$. We will use
the previous framework determining the parameters $p_1,\,p_2,\,p_3$ as
functions of the new parameters. We set
$$p_1+p_3=\pi_1 \qquad \mbox{and} \qquad p_2+p_3=\pi_2.$$
Then
$$r={p_3p_4-p_1p_2\over\sqrt{\pi_1(1-\pi_1)\pi_2(1-\pi_2)}}.$$
Then
$$p_3p_4-p_1p_2=r\sqrt{\pi_1(1-\pi_1)\pi_2(1-\pi_2)}.$$
Inserting here $p_4=1-p_1-p_2-p_3$, $p_1=\pi_1-p_3$ and $p_2=\pi_2-p_3$
we get
$$p_3=\pi_1\pi_2+ r\sqrt{\pi_1(1-\pi_1)\pi_2(1-\pi_2)}.$$
Then
\begin{equation}
\label{eq:piuno}
p_1=\pi_1-\pi_1\pi_2- r\sqrt{\pi_1(1-\pi_1)\pi_2(1-\pi_2)},
\end{equation}
\begin{equation}
\label{eq:pidue}
p_2=\pi_2-\pi_1\pi_2- r\sqrt{\pi_1(1-\pi_1)\pi_2(1-\pi_2)}.
\end{equation}
The only problem is that we have to make sure that both $p_1$ and $p_2$
turn out to be positive. So we have to impose
\begin{equation}
\label{eq:piunopositivo}
\pi_1(1-\pi_2)> r\sqrt{\pi_1(1-\pi_1)\pi_2(1-\pi_2)},
\end{equation}
and
\begin{equation}
\label{eq:piduepositivo}
\pi_2(1-\pi_1)> r\sqrt{\pi_1(1-\pi_1)\pi_2(1-\pi_2)}.
\end{equation}
This impose restrictions on the attainable upper bound for $r$. Indeed
it has to be
$$r<\min\left ({\pi_1(1-\pi_2)\over \sqrt{\pi_1(1-\pi_1)\pi_2(1-\pi_2)}},\,
{\pi_2(1-\pi_1)\over \sqrt{\pi_1(1-\pi_1)\pi_2(1-\pi_2)}}\right ).$$
Write $\pi_1=\beta\pi_2,\;\beta >0$. Then the above equation becomes
$$r<\min\left (\sqrt{{\beta (1-\pi_2)\over 1-\beta\pi_2}},\,
\sqrt{{1-\beta\pi_2\over \beta (1-\pi_2)}}\right ).$$
If $\beta>1$ the minimum is
$$\sqrt{{1-\beta\pi_2\over \beta (1-\pi_2)}}.$$
If $\beta<1$ the minimum is
$$\sqrt{{\beta (1-\pi_2)\over 1-\beta\pi_2}}.$$
If $\beta =1$ the minimum is 1. So when $Y_1$ and $Y_2$ are identically
distributed there is no constraint to impose on $r$.

\noindent
Now,
when $\beta>1$, if we take the derivative of the minimum with
respect to $\beta$, if we write
$$f(\beta)= {1-\beta\pi_2\over \beta (1-\pi_2)},$$
we have
$$f'(\beta)=-{1-\pi_2\over\beta^2(1-\pi_2)^2}<0,$$
so that the derivative of the minimum is negative: the minimum is a
decreasing function of $\beta$.

\noindent
When $\beta<1$, if we take the derivative of the minimum with
respect to $\beta$, if we write
$$f(\beta)= {\beta (1-\pi_2)\over 1-\beta\pi_2},$$
we have
$$f'(\beta)={(1-\pi_2)(1+\pi_2(1-\beta))\over\beta^2(1-\pi_2)^2}>0,$$
so that the derivative of the minimum is positive: the minimum is an
increasing function of $\beta$.

\noindent
In Table 1 we present the upper bound of $r$ for some couples
of values of $\pi_1$ and $\pi_2$. Let us note that this upper bound is the
same if we interchange $\pi_1$ with $\pi_2$ and if we substitute
$\pi_1$ with $1-\pi_1$ and $\pi_2$ with $1-\pi_2$.

\section{Regression Function}
Let us recall two consequences of the binomial theorem:
\begin{equation}
\label{eq:conseguenzauno}
\sum_{i=0}^ni{n\choose i}a^ib^{n-i}=na(a+b)^{n-1}.
\end{equation}
\begin{equation}
\label{eq:conseguenzadue}
\sum_{i=0}^ni^2{n\choose i}a^ib^{n-i}=na(a+b)^{n-2}[a+b+(n-1)a].
\end{equation}

\noindent
Because of Equation~\ref{eq:fondamentale} the joint probability function of
$X_1$ and $X_2+X_3$ is given by
$$\prob{X_1=k,\,X_2+X_3=h,\,X_4=n-k-h}=
{n!p_1^kp_4^{n-k-h}
\over k!(n-k-h)!h!}(p_2+p_3)^h.$$
Because of Equation~\ref{eq:fondamentaledue} the conditional distribution
of $X_1$ given $X_2+X_3$ is
$$\prob{X_1=k\vert X_2+X_3=h}=
{p_1^kp_4^{n-k-h}(n-h)!
\over k!(n-k-h)!(1-p_2-p_3)^{n-h}}.$$
Then the conditional expectation of $X_1$ given $X_2+X_3=h$ is given by,
noticing that the above conditioning event implies $0\le X_1\le n-h$,
\begin{eqnarray*}
\b{E}(X_1\vert X_2+X_3=h)&=&{1\over (1-p_2-p_3)^{n-h}}
\sum_{k=0}^{n-h}k{n-h\choose k}p_1^kp_4^{n-k-h}\\
&=&{1\over (1-p_2-p_3)^{n-h}}(n-h)p_1(p_1+p_4)^{n-h-1},
\end{eqnarray*}
where we used Equation~\ref{eq:conseguenzauno}. Now, since $p_1+p_4=
1-p_1-p_2$, we get
$$\b{E}(X_1\vert X_2+X_3=h)={p_1\over 1-p_2-p_3}(n-h).$$
Finally
the conditional expectation of $X_1$ given $X_2+X_3$ is
\begin{equation}
\b{E}(X_1\vert X_2+X_3)={p_1\over 1-p_2-p_3}(n-X_2-X_3).
\end{equation}
Along the same lines we evaluate now the joint probability function of
$X_3$ and $X_2+X_3$. We obtain
\begin{eqnarray*}
\prob{X_3=h,\,X_2+X_3=k}
&=&\sum_{i=0}^{n-k}\prob{X_1=i,\,X_2=k-h,\,
X_3=h,\,X_4=n-k-i}\\
&=&
\sum_{i=0}^{n-k}
{n!\over i!(k-h)!h!(n-k-i)!}p_1^ip_2^{k-h}
p_3^hp_4^{n-k-i}\\
&=&
{n!p_2^{k-h}p_3^h\over (k-h)!
h!}\sum_{i=0}^{n-k}
{i\over i!(n-k-i)!}p_1^ip_4^{n-k-i}\\
&=&
{k!n!p_2^{k-h}p_3^h\over k!(k-h)!
h!}\sum_{i=0}^{n-k}
{i\over i!(n-k-i)!}p_1^ip_4^{n-k-i}\\
&=&{k\choose h}{n\choose k}p_2^{k-h}p_3^h(p_1+p_4)^{n-k}.
\end{eqnarray*}
Then the conditional distribution
of $X_3$ given $X_2+X_3$ is
$$\prob{X_3=h\vert X_2+X_3=k}=
{k\choose h}{p_2^{k-h}p_3^h\over (p_2+p_3)^k}.$$
Hence the conditional expectation of $X_3$ given $X_2+X_3=k$ is given by,
noticing that the above conditioning event implies $0\le X_3\le k$,
\begin{eqnarray*}
\b{E}(X_3\vert X_2+X_3=k)&=&
\sum_{h=0}^kh{k\choose h}{p_2^{k-h}p_3^h\over (p_2+p_3)^k}\\
&=&{p_2^k\over (p_2+p_3)^k}
\sum_{h=0}^kh{k\choose h}\left ({p_3\over p_2}\right )^h\\
&=&{p_2^k\over (p_2+p_3)^k}
k {p_3\over p_2}\left (1+{p_3\over p_2}\right )^{k-1}\\
&=&{kp_3\over p_2+p_3},
\end{eqnarray*}
where we used again Equation~\ref{eq:conseguenzauno}.
Finally
the conditional expectation of $X_3$ given $X_2+X_3$ is
\begin{equation}
\b{E}(X_3\vert X_2+X_3)={p_3\over p_2+p_3}(X_2+X_3).
\end{equation}
It follows
\begin{eqnarray}
\b{E}(Y_1\vert Y_2)&=&\b{E}(X_1+X_3\vert X_2+X_3)\nonumber\\
&=&\b{E}(X_1\vert X_2+X_3)
+\b{E}(X_3\vert X_2+X_3)\nonumber\\
&=&{p_1\over 1-p_2-p_3}(n-X_2-X_3)
+{p_3\over p_2+p_3}(X_2+X_3)\nonumber\\
&=&\alpha n+(\beta-\alpha)Y_2,
\end{eqnarray}
where we wrote $\alpha = {p_1\over 1-p_2-p_3}$ and
$\beta = {p_3\over p_2+p_3}$.
We can conclude that the regression function is linear.

\section{Conditional Variance}
Since
\begin{eqnarray*}
&&\prob{X_1=h,\,X_2+X_3=k,\,X_3=r,\,X_4=n-k-h}=\\
&&\qquad\qquad ={n!\over h!(k-r)!r!(n-k-h)!}p_1^hp_2^{k-r}p_3^rp_4^{n-k-h},
\end{eqnarray*}
the joint conditional probability function of $X_1$ and $X_3$ given
$X_2+X_3$ is obtained as
$$\prob{X_1=h,\,X_3=r\vert X_2+X_3=k}={k\choose r}{n-k\choose h}
{p_1^hp_2^{k-r}p_3^rp_4^{n-k-h}\over (p_2+p_3)^k(1-p_2-p_3)^{n-k}}.$$
Since
$$\prob{X_1=h\vert X_2+X_3=k}={n-k\choose h}
{p_1^hp_4^{n-k-h}\over (1-p_2-p_3)^{n-k}},$$
and
$$\prob{X_3=r\vert X_2+X_3=k}=
{k\choose r}
{p_2^{k-r}p_3^r\over (p_2+p_3)^k},$$
we see that $X_1$ and $X_3$ are conditionally independent given $X_2+X_3$.

\noindent
It follows that
$$\b{V}ar(X_1+X_3\vert X_2+X_3)=
\b{V}ar(X_1\vert X_2+X_3)+\b{V}ar(X_3\vert X_2+X_3).$$
We obtain
\begin{eqnarray*}
\b{E}(X_1^2\vert X_2+X_3=h)&=&{1\over (1-p_2-p_3)^{n-h}}
\sum_{k=0}^{n-h}k^2{n-h\choose k}p_1^kp_4^{n-k-h}\\
&=&{1\over (1-p_2-p_3)^{n-h}}(n-h)p_1(p_1+p_4)^{n-h-2}\\
&&\qquad\times
[p_1+p_4+(n-h-1)p_1]\\
&=&
{1\over (p_1+p_4)^2}(n-h)p_1
[p_1+p_4+(n-h-1)p_1],
\end{eqnarray*}
where we used Equation~\ref{eq:conseguenzadue} and the fact that $1-p_2-p_3=
p_1+p_4$. Thus
$$\b{V}ar(X_1\vert X_2+X_3=h)={p_1p_4\over (p_1+p_4)^2}(n-h),$$
that is
$$\b{V}ar(X_1\vert Y_2)={p_1p_4\over (p_1+p_4)^2}(n-Y_2).$$
Analogously
\begin{eqnarray*}
\b{E}(X_3^2\vert X_2+X_3=h)&=&{p_2^h\over (p_2+p_3)^h}
\sum_{k=0}^hk^2{h\choose k}\left ({p_3\over p_2}\right )^k\\
&=&{p_2^h\over (p_2+p_3)^h}h\left ({p_3\over p_2}\right )
\left (1+{p_3\over p_2}\right )^{h-2}\\
&&\qquad\times
\left [1+ {p_3\over p_2}+(h-1)\left ({p_3\over p_2}\right )\right ]\\
&=&
{hp_3\over (p_2+p_3)^2}
[p_2+p_3+(h-1)p_3],
\end{eqnarray*}
where we used Equation~\ref{eq:conseguenzadue}.
It follows
$$\b{V}ar(X_3\vert X_2+X_3=h)={hp_2p_3\over (p_2+p_3)^2},$$
that is
$$\b{V}ar(X_3\vert Y_2)={p_2p_3\over (p_2+p_3)^2}Y_2.$$
Finally
$$\b{V}ar(Y_1\vert Y_2)=\gamma +\delta Y_2,$$
where we set $\gamma = {np_1p_4\over (p_1+p_4)^2}$ and
$\delta ={p_2p_3\over (p_2+p_3)^2}-{p_1p_4\over (p_1+p_4)^2}$.
We can conclude that the regression function is linear but not
homoscedastic.

\bigskip
\begin{center}
{\em  Table 1. Upper bound of $r$ for selected values of
$\pi_1$ and $\pi_2$.}

\begin{tabular}{||c|c|c|c|c|c|c|c|c|c||}\hline\hline
& \multicolumn{9}{c||}{$\pi_2$}\\ \cline{2-10}
$\pi_1$&0.1&0.2&0.3&0.4&0.5&0.6&0.7&0.8&0.9\\ \hline
0.1&1&0.667&0.509&0.409&0.333&0.272&0.218&0.167&0.111\\
\hline
0.2&0.667&1&0.767&0.612&0.500&0.408&0.327&0.250&0.167\\
\hline
0.3&0.509&0.767&1&0.802&0.655&0.534&0.428&0.327&0.218\\
\hline
0.4&0.409&0.612&0.802&1&0.816&0.667&0.534&0.408&0.272\\
\hline
0.5&0.333&0.500&0.655&0.816&1&0.816&0.655&0.500&0.333\\
\hline
0.6&0.272&0.409&0.534&0.667&0.816&1&0.802&0.612&0.409\\
\hline
0.7&0.218&0.327&0.428&0.534&0.655&0.802&1&0.767&0.509\\
\hline
0.8&0.167&0.250&0.327&0.409&0.500&0.612&0.767&1&0.667\\
\hline
0.9&0.111&0.167&0.218&0.272&0.333&0.409&0.509&0.667&1\\
\hline
\end{tabular}
\end{center}

\end{document}